\documentclass[twocolumn,prb,superscriptaddress, showpacs]{revtex4}
\usepackage[final]{graphicx}
\usepackage[usenames]{color}
\usepackage{afterpage}
\begin{document}

\title{First-principles investigation of the electron-phonon interaction in OsN$_2$:   Theoretical prediction of superconductivity mediated by N-N covalent bonds}

\author{Alexander D. Hern\'andez}
\affiliation{Centro At\'{o}mico Bariloche, 8400 San Carlos de Bariloche, Argentina}
\affiliation{The Abdus Salam International Centre for Theoretical Physics, Strada
Costiera 11, 34014 Trieste, Italy}
\author{Javier A. Montoya}
\affiliation{INFM/Democritos National Simulation Center, via Beirut
2-4, 34014 Trieste, Italy} 
\affiliation{SISSA--International School for Advanced Studies, via Beirut 2-4,
34014 Trieste, Italy}
\author{Gianni Profeta}
\affiliation{CNISM - Dipartimento di Fisica,
Universit\`a degli Studi dell'Aquila, Via Vetoio 10,
I-67010 Coppito (L'Aquila) Italy}
\author{Sandro Scandolo}
\affiliation{The Abdus Salam International Centre for Theoretical Physics, Strada
Costiera 11, 34014 Trieste, Italy}
\affiliation{INFM/Democritos National Simulation Center, via Beirut
2-4, 34014 Trieste, Italy} 

\date{\today}

\begin{abstract}
A first-principles investigation of the electron-phonon interaction in the recently synthesized
osmium dinitride (OsN$_2$) compound predicts that the material is a superconductor.
Superconductivity in OsN$_2$ would originate from the stretching of covalently bonded dinitrogen units embedded in the transition-metal matrix, thus adding dinitrides to the class of superconductors containing covalently bonded light elements. The dinitrogen vibrations are strongly coupled to the electronic states at the Fermi level and generate narrow peaks in the Eliashberg spectral function $\alpha^2F(\omega)$. The total electron-phonon coupling of OsN$_2$ is $\lambda=0.37$ and the estimated superconducting temperature T$_c \approx 1$ K. We suggest that the superconducting temperature can be substantially increased by hole doping of the pristine compound and show that T$_c$ increases to 4 K with a doping concentration of 0.25 holes/OsN$_2$ unit.
\end{abstract}


\maketitle

Metallic compounds containing light elements such as H, Li, and B
have attracted considerable attention recently due to their 
potential superconducting properties. \cite{ashcroft:2004, shimizu:2002, eremets:2001}
Within the weak-coupling BCS theory, high-frequency phonons due to the presence of
atoms with light masses ensure a large prefactor in the BCS formula for the
superconducting critical temperature $T_c$. Thus, even a moderate 
electron-phonon coupling can yield a sizable T$_c$. \cite{ashcroft:1997, kortus:2001}
The discovery of superconductivity in MgB$_2$, \cite{nagamatsu:2001} B-doped diamond, \cite{ekimov:2004} B-doped silicon, \cite{bustarret:2006} and calcium and ytterbium graphite intercalated compounds \cite{weller:2005} confirms this picture and extends it by showing that strong covalent bonds between light atoms can provide a sizable contribution to the electronic density of states at the Fermi level, under appropriate ``doping'' conditions. In MgB$_2$, the Fermi level crosses the covalent $\sigma$ 
bonds formed by boron atoms. Such states are partially empty as a consequence of the 
lowering of the $\pi$ bands, caused by the Mg$^{2+}$ attractive potential felt by the B-$\pi$ electrons. \cite{nagamatsu:2001, kortus:2001} In B-doped diamond, \cite{ekimov:2004} substitutional boron atoms provide hole doping to the C-C $sp^3$ covalent bonds. The
strong C-C bonding allows the structure to remain stable even at high doping. In electron-doped graphite intercalated compounds, the Fermi level crosses the graphitic C-$\pi$ band and the intercalated band.

Nitrogen follows boron and carbon in the first row of the Periodic Table and is characterized, in its elemental form, by a strong triple bond in the low pressure molecular phases and by covalent single bonds in the nonmolecular phase stable at pressures exceeding a megabar. \cite{eremets:2004} As a consequence, molecular phases are insulating and the nonmolecular phase is semiconducting. \cite{eremets:2001b} 
In analogy with the boron and carbon-based superconducting compounds described above, search 
for superconductivity in nitrogen-based systems requires the identification of compounds where 
covalent bonds between nitrogen atoms persist in a stable form in the presence of doping species
and of a resulting metallic state. 

\begin{figure}[htp]
\includegraphics[width=1.0\linewidth]{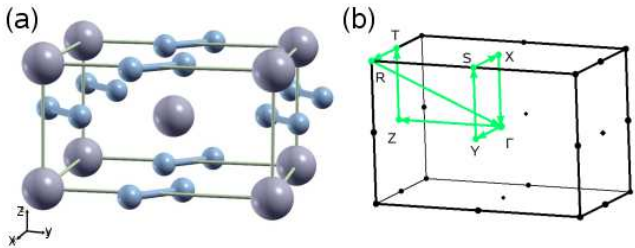}
\caption{ (Color online) (a) Crystal structure of OsN$_2$ (isostructural to marcasite). The space group is {\it Pnnm}, with osmium atoms (gray) in the Wyckoff sites $2a$ and nitrogen atoms (blue) in the $4g$ sites (Ref. \onlinecite{montoya:2007}). (b) Brillouin zone showing the high-symmetry directions used in Figs.~\ref{Bands} and \ref{Ph_disp}.}
\label{FermiSurface}
\end{figure}

To our knowledge only OsN$_2$, a member of the family of late transition-metal 
nitrides synthesized recently at high pressure and temperature starting from their constituent elements, \cite{gregoryanz:2004,young:2006_1,Goncharov:2006,young:2006_2} fulfills the above criteria. The compounds have been obtained by subjecting the parent metal to extreme conditions of pressure and temperature in a nitrogen embedding medium, in a diamond-anvil cell. Interest in these compounds has resided so far in their large bulk modulus, which suggests superhard mechanical properties. {\it Ab initio} calculations show that among the three compounds synthesized so far (PtN$_2$, IrN$_2$, and OsN$_2$) only OsN$_2$ has a metallic character,  \cite{montoya:2007,angewandte} in agreement with the experimentally observed absence of first-order Raman peaks in this compound. \cite{young:2006_1} Covalently bonded dinitrogen (N$_2$) units are preserved in the marcasitelike crystal structure of OsN$_2$ (Ref. \onlinecite{montoya:2007}) [see Fig.~\ref{FermiSurface}(a)], which makes 
this compound an obvious candidate to investigate superconductivity in nitrogen-rich systems.

In this work, we investigate the superconducting properties of
OsN$_2$ in the framework of a phonon mediated pairing mechanism. 
We performed {\it ab initio} calculations of the
Fermi surface, electronic bands, phonon dispersions, and
electron-phonon couplings in OsN$_2$, and demonstrate that the
high-frequency modes originating from the covalently bonded N$_2$ units
are strongly coupled to the electronic states at the Fermi level and would give rise 
to a T$_c$ of about 1 K. We also show that the superconducting temperature can be 
greatly increased by hole doping.

The calculations were performed with the QUANTUM ESPRESSO package \cite{QE} employing density functional theory and the Perdew-Burke-Ernzerhoff exchange-correlation functional. \cite{pbe} 
An ultrasoft pseudopotential description of the ion-electron interaction, \cite{vanderbilt} 
with Os $5s$ and $5p$ semicore electrons included in the valence, was used together with a plane-wave 
basis set for the electronic wave functions and the charge density, with energy cutoffs of 40 and 480 Ry, respectively. The dynamical matrices and the electron-phonon coupling constants 
$\lambda$ were calculated using density functional perturbation theory (DFPT) in the linear 
response regime. \cite{QE,baroni} The electronic Brillouin zone (BZ) integration in the
phonon calculation was sampled with a $20\times 16\times 30$ uniform k-point mesh.
The electron-phonon coupling was found to be converged with a
finer grid of $26\times 22\times 40$ k points and a Gaussian smearing of 0.006 Ry.
The dynamical matrix was computed on a $2^3$ mesh of phonon wave vectors q.
The phonon dispersion was then obtained on a finer $8^3$ q mesh
by Fourier interpolation of the real space interatomic force constants.
In this way, $\lambda$ is calculated over a $8^3$ q-point mesh.

\begin{figure}[htp]
\includegraphics[width=1.0\linewidth]{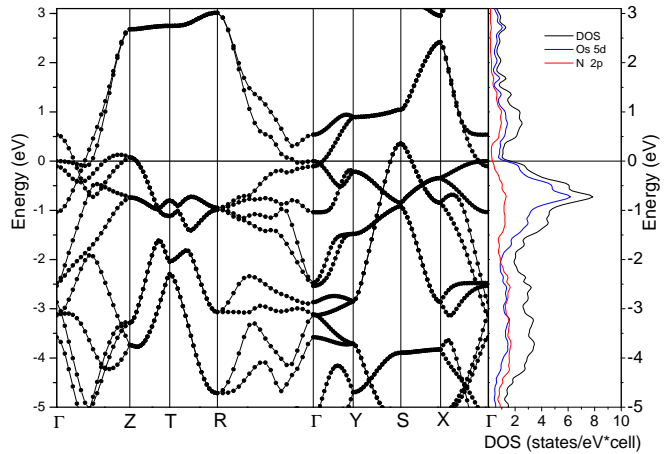}
\caption{ (Color online) Left panel: electronic bands of marcasite OsN$_2$ at ambient
pressure along the high-symmetry lines shown in Fig.~\ref{FermiSurface}(b). Right panel:
electronic density of states and its projection onto the Os $5d$ (blue) and N $2p$ (red) 
orbitals. The Fermi energy is set to zero.}
\label{Bands}
\end{figure}

In the energy range shown in Fig.~\ref{Bands}, the total electronic density of states (DOS) 
is essentially determined 
by Os $5d$ and N $2p$ orbitals (Fig.~\ref{Bands}, right panel). At the Fermi level, the N $2p$ orbitals contribute
with about 20\% of the total DOS. The projection on the atomic orbitals
also shows that 92\% of the nitrogen contribution at E$_F$ is due to the 
N p$_{x,y}$ states and the remaining 8\% to N p$_z$ states. 
The N p$_{x,y}$ orbitals lie in the plane containing the N-N units, and are thus 
directly involved in the formation of the N-N covalent bond in OsN$_2$. 
Integrating the DOS in a window of energy close to E$_F$
(between E$_F$ and 1 eV below E$_F$), we find an antibonding character for the electronic states on the N-N units. This is consistent with the considerable weakening of the N-N bond in OsN$_2$ with respect to the molecular triple bond, and is confirmed by the large reduction of the N-N stretching 
frequency from 2300 cm$^{-1}$ in the molecular state to 500-800 cm$^{-1}$ in the compound. 
A non-negligible coupling of the electronic 
states close to E$_F$ with the N-N stretching vibrational modes can thus be anticipated based on simple band-structure considerations. 
It is interesting to remark that the presence of the Os framework is crucial to the presence of a finite nitrogen component in the DOS at E$_F$. A band-structure calculation for a pure nitrogen system obtained by removing the Os atoms from the crystal structure of OsN$_2$ gives an insulating solid, indicating that the nitrogen component at E$_F$ arises from the coupling of the N-N units with the transition-metal framework, through rehybridization and/or charge transfer.

\begin{figure}[htp]
\includegraphics[width=1.0\linewidth]{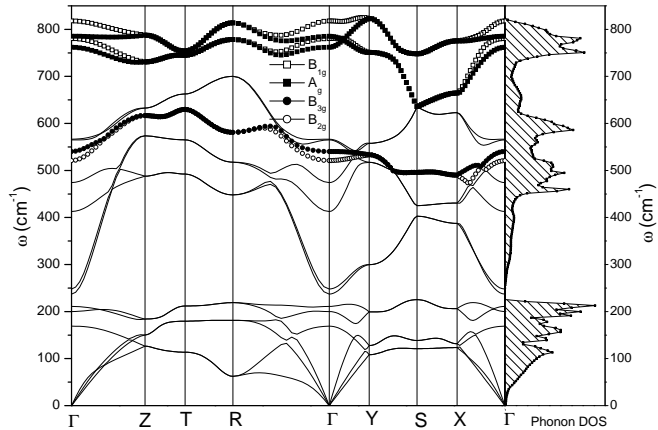}
\caption{\label{Ph_disp} Phonon dispersions along high-symmetry directions of the Brillouin zone and corresponding phonon density of states (right panel).}
\end{figure}

Due to the presence of stiff N-N bonds, the calculated phonons of OsN$_2$, 
shown in Fig.~\ref{Ph_disp}, can be divided into 
three main groups: a low-frequency group (up to 200 cm$^{-1}$) involving mostly the Os 
sublattice, an intermediate group corresponding to the librational modes of the N-N units
(between 250 and 600 cm$^{-1}$), and a high-frequency manifold corresponding to rotation and 
stretching of the N-N units (above 600 cm$^{-1}$). The 18 phonon modes of the marcasite 
structure belong to eight irreducible representations.
Four of these representations (B$_{1g}$, B$_{2g}$, B$_{3g}$, and A$_g$) 
are associated with nitrogen displacements only, with osmium atoms at rest,
and are highlighted with symbols in Fig.~\ref{Ph_disp}. Representations B$_{2g}$ and B$_{3g}$ 
have one mode each and involve dinitrogen vibrations along the $\hat{z}$ axis, 
while representations A$_g$ and B$_{1g}$ have two modes and involve vibrations along the 
$\hat{x}$ and $\hat{y}$ directions. In B$_{2g}$ and A$_g$ both dinitrogen units oscillate in phase, while in B$_3g$ and B$_{1g}$ the N-N units vibrate in counterphase. As we can see from Fig.~\ref{Ph_disp}, the nitrogen light mass and the covalent N-N bond ensure a high frequency for these modes, 
with A$_g$ and B$_{1g}$ modes ranging between 640 and 825 cm$^{-1}$. Raman peaks in this frequency range 
have been observed experimentally also in PtN$_2$ and IrN$_2$ and have been associated with the stretching
of N-N units. \cite{gregoryanz:2004,young:2006_1,Goncharov:2006}
As a confirmation of the planar ($xy$) nature of bonding in N-N, we note that the $xy$-polarized phonons in 
the A$_g$ and B$_{1g}$ representations are higher in frequency with respect to $z$-polarized modes.

Within DFPT, \cite{baroni} the electron-phonon interaction for a
phonon mode $\nu$ with momentum ${\bf q}$ can be calculated as
\begin{eqnarray}\label{elph}
\lambda_{{\bf q}\nu} = \frac{4}{\omega_{{\bf q}\nu}N(E_F) N_{k}} \sum_{{\bf k},n,m}
|g_{{\bf k}n,{\bf k+q}m}^{\nu}|^2 \delta(\epsilon_{{\bf k}n}) \delta(\epsilon_{{\bf k+q}m}) \nonumber
\end{eqnarray}
where the sum is over the Brillouin zone. The matrix element is
$g_{{\bf k}n,{\bf k+q}m}^{\nu}= \langle {\bf k}n|\delta V/\delta u_{{\bf q}\nu} |{\bf k+q} m\rangle /\sqrt{2 \omega_{{\bf q}\nu}}$,
where $u_{{\bf q}\nu}$ is the amplitude of the displacement of the phonon
and $V$ is the Kohn-Sham potential.
The electron-phonon coupling is calculated as a BZ average
over the phonon wave vectors
$\lambda=\sum_{{\bf q}\nu} \lambda_{{\bf q}\nu}/N_q$.
The Eliashberg spectral function $\alpha^2F(\omega)$ is defined as
\begin{equation}
\alpha^2F(\omega)=\frac{1}{2 N_q}\sum_{{\bf q}\nu}
\lambda_{{\bf q}\nu} \omega_{{\bf q}\nu} \delta(\omega-\omega_{{\bf q}\nu} )
\label{Eqa2F}
\end{equation}
and allows to compute
$\lambda(\omega)=2 \int_{0}^{\omega} d\omega^{\prime}
\alpha^2F(\omega^{\prime})/\omega^{\prime}$.

\begin{figure}[htb]
\includegraphics[width=1.0\linewidth]{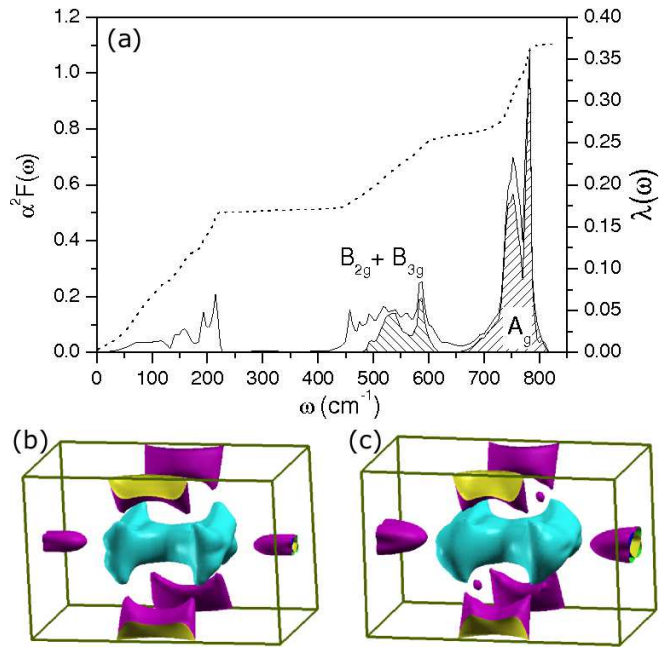}
\caption{(Color online) (a) Eliashberg function $\alpha^2F(\omega)$ (continuous line) and integrated coupling 
$\lambda(\omega)$ (dashed line) of OsN$_2$. The shaded regions are the B$_{2g}$, B$_{3g}$, and A$_g$ 
contributions to $\alpha^2F$. (b) Calculated Fermi surface of OsN$_2$. The Fermi surface consists of 
three electron pockets located close to the zone center, only one of which is visible (green pocket), and 
of four hole pockets, two of them centered at Z and two at the S point. (c) Fermi surface upon distortion 
of the lattice along an A$_g$ phonon.}
\label{a2F} 
\end{figure}

Figure~\ref{a2F}(a) shows the Eliashberg spectral function $\alpha^2F(\omega)$ calculated from Eq.~(\ref{Eqa2F}).
Three separate contributions to the electron-phonon interaction can be distinguished and attributed to the low-frequency, intermediate-frequency, and high-frequency phonons, respectively.
Two high-frequency peaks are particularly strong and well resolved, and are associated with 
the contribution of the two A$_g$ phonons, while the contribution of the B$_{2g}$ and B$_{3g}$ phonons 
accounts for most of the electron-phonon interaction in the intermediate-frequency range from 500 to 600 cm$^{-1}$. The integral $\lambda(\omega)$, represented by a dashed line in Fig.~\ref{a2F}(a), shows that the low-frequency phonons that involve mostly osmium atoms account for a contribution of $\lambda=0.17$. The high-frequency phonons associated with the stretching of the covalently bonded N-N units contribute with an 
equivalent amount, which brings the total $\lambda$ for OsN$_2$ to 0.37. 
More insight about the nature of the electron-phonon interaction that leads to such a large contribution to $\lambda$ from N-N bonds can be obtained by analyzing the changes of the Fermi surface (FS) that arise upon 
distorting the lattice along the relevant phonon modes. In particular, we concentrate on the A$_g$ 
modes that show the highest $\alpha^2F(\omega)$ values and dominate the high-frequency contributions. 
In Fig.~\ref{a2F}(b), we compare the FS of the undistorted OsN$_2$ with that of a distorted OsN$_2$ crystal 
obtained by changing by 4\% the distance between nitrogen atoms in the N-N units along the 
$xy$-polarized A$_g$ mode. The most relevant change in the FS is the migration of electrons from the bands that contain the hole pockets at Z to the bands that contain the electron pockets near $\Gamma$. 
Interband electron transfer is generally associated with a large
electron-phonon coupling, which is consistent with our finding of a
large contribution of the A$_g$ modes to $\lambda$.

The superconducting critical temperature can be estimated using the
McMillan formula: \cite{mcmillan}
\begin{equation}
T_c = \frac{\langle \omega_{\mathrm{ln}} \rangle}{1.2} \exp\left( - \frac{1.04 (1+\lambda)}{\lambda-\mu^* (1+0.62\lambda)}\right), \label{eq:mcmillan}
\end{equation}
where $\mu^*$ is the screened Coulomb pseudopotential and
$\langle\omega_{{\mathrm{ln}}}\rangle=280$ K is the logarithmically
averaged phonon frequency. We obtain 0.3$<$T$_c$$<$1.2 for  0.13$>\mu^*$$>$0.08 
(with T$_c \approx 1$ K at the widely accepted value $\mu^*$=0.1, 
see Refs. \onlinecite{mcmillan,morel:1962,comment}). 

A careful examination of the electronic DOS of OsN$_2$ (Fig.~\ref{Bands}) suggests that hole doping could further enhance T$_c$. Hole doping in OsN$_2$ would, in fact, lower the Fermi level toward a region of higher electronic DOS and would, at the same time, stiffen the N-N bonds by partially emptying the antibonding states below E$_F$.
The reported synthesis of several nitrides from different transition metals (Pt, Ir, Os, Pd) suggests that the synthesis of transition-metal nitride {\em alloys}, i.e., of compounds with N-N units inserted in a matrix of mixed metal composition, is not impossible. Alloys with different composition allow a tuning of the  electronic DOS, as observed in transition-metal alloys. \cite{chu:1971}

\begin{figure}[htb]
\includegraphics[width=1.0\linewidth]{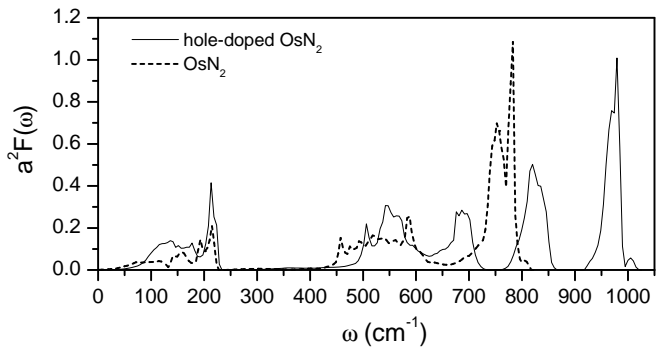}
\caption{\label{a2Fholedoped} Eliashberg function $\alpha^2F(\omega)$ of
hole-doped OsN$_2$ (continuous line) and undoped OsN$_2$ (dashed
line).}
\end{figure} 

In order to explore the consequences of hole doping, we carried out {\it ab initio} calculations of OsN$_2$ 
with a hole doping of 0.5 holes/unit cell, corresponding, e.g.,  to a hypothetical alloy with Os$_{0.75}$Re$_{0.25}$N$_2$ 
composition. As expected, the DOS at E$_F$ increases about 2.4 times with respect to the undoped case. The electron-phonon coupling matrix elements remain approximately the same in the doped and undoped cases, but the phonon frequencies associated with the N-N modes increase by about 
200 cm$^{-1}$ (see Fig.~\ref{a2Fholedoped}), which confirms the strong coupling of these 
modes with the electronic states close to the Fermi level. The frequency increase causes 
an increase of $\langle\omega_{{\mathrm{ln}}}\rangle$ to $310$ K. The total electron-phonon coupling 
parameter, $\lambda$, increases to 0.49, leading to a superconducting critical temperature of $\simeq$ 4K for doped OsN$_2$.

In conclusion, we predict that OsN$_2$ is a superconductor, and that
its superconducting properties are connected with a strong coupling
between the stretching modes of the covalently bonded N$_2$ units with
the electronic states at the Fermi level, similar to what has been observed in
a number of boron and carbon-based compounds, including MgB$_2$. We
predict an enhancement of the superconducting temperature by doping
OsN$_2$ with holes, which we believe can be achieved experimentally by
synthesizing the nitride starting from a hole-doped Os alloy. We hope
this work will stimulate the experimental search for other members of the
dinitride family with metallic character and potential superconducting properties.

We acknowledge useful conversations with E. Gregoryanz and R. Rousseau. A.D.H. thank C. Proetto and D. Dominguez for useful suggestions and support from CONICET (PIP 5596) and ANPCyT (PICT 13829 and PICT 13511). The OsN$_2$ crystal structure and Fermi surface was generated with A. Kokalj's XCRYSDEN package ({\tt http://www.xcrysden.org/}).


\begin{thebibliography}{99}
\bibitem{comment} {In order to validate our prediction 
of a possible superconducting phase, we performed preliminary calculations 
of the critical temperature within the superconducting density functional 
theory (SCDFT) (Ref. \onlinecite{SCDFTI}). Within the isotropic approximation and the Thomas-Fermi approximation for the screening of the electron-electron Coulomb interaction (Ref. \onlinecite{SCDFTII}) the calculated T$_c$ is 0.94 K. Multiband superconductivity and anisotropy of the electron-phonon interaction could increase the critical temperature, and will be considered in a 
 dedicated work.}
\end{thebibliography}
\end{document}